# Distributed Inference and Query Processing for RFID Tracking and Monitoring [*]

Zhao Cao, Charles Sutton[†], Yanlei Diao, Prashant Shenoy
Department of Computer Science    [†]School of Informatics
University of Massachusetts, Amherst    University of Edinburgh
caozhao, yanlei, shenoy@cs.umass.edu, [†]csutton@inf.ed.ac.uk

## ABSTRACT

In this paper, we present the design of a scalable, distributed stream processing system for RFID tracking and monitoring. Since RFID data lacks containment and location information that is key to query processing, we propose to combine location and containment inference with stream query processing in a single architecture, with inference as an enabling mechanism for high-level query processing. We further consider challenges in instantiating such a system in large distributed settings and design techniques for distributed inference and query processing. Our experimental results, using both real-world data and large synthetic traces, demonstrate the accuracy, efficiency, and scalability of our proposed techniques.

## 1. INTRODUCTION

RFID is a promising electronic identification technology that enables a real-time information infrastructure to provide timely, high-value content to monitoring and tracking applications. An RFID-enabled information infrastructure is likely to revolutionize areas such as supply chain management, healthcare, and pharmaceuticals [9]. Consider, for example, a healthcare environment such as a large hospital that tags all pieces of medical equipment (e.g., scalpels, thermometers) and drug products for inventory management. Each storage area or patient room is equipped with RFID readers that scan medical devices, drug products, and their associated cases. Such an RFID-based infrastructure offers a hospital unprecedented near real-time ability to track and monitor objects and detect anomalies (e.g., misplaced objects) as they occur. The use of RFID tags provide similar benefits in distributed supply chains where objects, cases and pallets must be tracked, and in pharmaceutical environments that require combating counterfeit drugs and preventing pilfering. To illustrate, consider the following types of continuous queries that may be posed on the RFID streams:

- *Tracking queries*, which include queries such as "report any pallet that has deviated from its intended path," or "list the path taken by a medical device equipment through the hospital before it was misplaced." Such tracking queries are *location* queries that require object locations or location histories.

- *Containment queries,* which include queries such as "raise an alert if a flammable item is not packed in a fireproof case," or "verify that food containing peanuts is never exposed to other food cases for more than an hour." This class of queries involve inter-object relationships, e.g., containment between objects, cases, and pallets, and are useful for enforcing packaging and shipping regulations.

- *Hybrid queries,* which include "for any temperature sensitive drug product, raise an alert if it has been placed outside a freezer and exposed to room temperature for 6 hours." This class of queries combine sensors streams (e.g., temperature) and RFID streams (e.g., object location and containment) to detect various conditions.

Unfortunately, the nature of RFID data makes these queries difficult to answer. The key challenge is that although such anomaly detection queries typically involve object locations and inter-object relationships such as containment, the RFID data does not directly contain this information. Rather, the data contains only the observed tag id and the reader id; this is a fundamental limitation of RFID technology. To enable queries on the data that is not actually available, the key is to exploit statistical regularities in the tag id and reader information so that one can *estimate* object locations and object relationships. The estimation problem is complex, however, because RFID readings are inherently noisy due to the sensitivity of radio frequency to occluding metal objects and interference [7]. For example, in our lab setup (Section 5.2), we observed read rates of 70%-85% even with state-of-the-art readers and tags. Real deployments in complex environments such as hospitals would be expected to experience similar issues.

A second key challenge is that environments such as large hospitals or supply chains are distributed in scope, for which a centralized approach may be limiting. In a centralized approach, RFID streams from various readers are sent to a central location for query processing. This approach can fail to scale because of the bandwidth overheads incurred due to high data volume and can also potentially increase the latency of detecting anomalous events, especially in geographically distributed settings. In contrast, a distributed approach processes data streams as they emerge, thereby reducing the delay of answering queries. However, as objects move from one location to another, tracking and monitoring queries must also "move" with these objects. To do so, both the state of objects and the state of monitoring queries relevant to these objects must be transferred to the new location to seed the computation there.

**Research contributions:** In this paper, we present the design of a scalable, distributed stream processing system for RFID tracking and monitoring. Our system combines location and containment

---

[*]This work has been supported in part by NSF grants IIS-0746939, IIS-0812347, and CNS-0923313.

inference with stream query processing into a single architecture, with inference as an enabling mechanism for high-level query processing for tracking, monitoring, and anomaly detection. We further scale such inference and thus enable query processing in large distributed environments that span multiple sites and numerous objects. More specifically, our contributions include the following:

***Novel statistical framework*** (Section 3). The key novelty in our approach to location and containment inference is to introduce the notion of *smoothing over object relations*, whereas all existing work on RFID data cleaning [8, 11] and location inference [14, 16] is limited to the traditional approach of smoothing over time. In contrast to temporal smoothing approaches, smoothing over containment in our work leads to a much simpler graphical model, thereby allowing more efficient inference. At the same time, our model and inference techniques can still accurately estimate location and containment information, so that high-level query processing can return high-quality answers.

Our general approach is as follows: (*i*) Our probabilistic model describes a physical world comprising object locations, containment relationships, and noisy RFID readings. (*ii*) We devise an inference algorithm, called RFINFER, for our model, working within an expectation maximization (EM) framework. The design of our model allows us to derive a simple customized M-step, which is essential for working at scale but still offers provable optimality. Furthermore, our algorithm is developed in an *unsupervised learning* framework; that is, it does not use machine learning techniques that require access to any specially-generated training data. (*iii*) We finally extend our algorithm to also detect changes of containment using a statistical method called *change point detection*.

***Distributed inference and query processing*** (Section 4). To suit the increasing scale of RFID tracking and monitoring, we develop a distributed approach that performs inference and query processing locally at each location, but transfers the state of inference and state of query processing as objects move across sites. A naive inference algorithm would incur high transfer overhead by requiring the entire history of observations collected from multiple sites over a long period of time. Instead, we propose to truncate history by sifting out the observations most informative about the true containment, and further distill such useful history into a few numbers for each object to minimize the inference state transferred. In distributed processing of tracking and monitoring queries, the main issue is that we need to transfer one copy of query state for each object. Our work exploits the inference results, in particular, stable containment to share query state among objects.

***Performance evaluation*** (Section 5). Our evaluation, using both real-world data and large synthetic traces, shows the following: (*i*) Our inference algorithm is highly accurate, with less than 7% error on containment and 0.5% error on location, for noisy traces with stable containment. (*ii*) With containment changes, our algorithm can achieve 85% accuracy when read rates reach 0.7 while keeping up with stream speed, as shown using real lab traces and simulations. (*iii*) Our distributed inference method offers 3 orders of magnitude reduction in communication cost over a centralized method without compromising accuracy, and scales to millions of objects over multiple sites. (*iv*) Our highly accurate inference allows a query processor to produce high-quality answers and further exploit sharing of query state across objects for state migration.

## 2. BACKGROUND

In this section, we provide background on RFID technology and RFID tracking and monitoring applications. Our system targets any distributed environment with multiple locations such as hospitals with multiple storage areas, supply chains with multiple warehouses, etc. For ease of exposition, the rest of this paper assumes that the environment is a distributed supply chain; however, our techniques are general and can be applied to other domains as well. Each item in the supply chain is assumed to be packed into a case, and multiple cases packed onto a pallet, which yields a containment relationship between items, cases and pallets. Items, cases, and pallets are assumed to be tagged. Each tag has a unique identity; the tag id can also indicate the level of packaging, e.g., a pallet, a case, or an item. We focus on passive RFID tags, which are battery-less and have a small amount of on-board memory, e.g., 4-64 KB. This memory is writable and can be exploited to store supply-chain-specific object state and enable "*querying anytime anywhere*".[1] We assume that each distribution center employs multiple RFID readers, for example, at the entry and exit points as well as at the belt and shelves to scan resident objects. Each such reader periodically interrogate tags in its read range and immediately returns the sensed data in the form of ($time$, $tag\_id$, $reader\_id$). The local servers of a distribution center collect raw RFID data streams from all readers and process these streams. The data streams from different centers are further aggregated to support global tracking and monitoring.

We next illustrate the tracking and monitoring queries that our work aims to support. Such queries assume an event stream with rich information including ($time$, $tag\_id$, $location$, $container$) and optional attributes describing object properties, such as the type of food or type of container (which can be obtained from the manufacturer's database). Note the different schemas for raw RFID readings and events used in query processing—events in the latter schema are produced by an inference module as we discuss shortly.

Query 1 is an example of a hybrid query that combines object locations, containment relationships, and temperature sensor readings. This query raises an alert if a frozen food or drug product has been placed outside a freezer and exposed to room temperature for 6 hours. The query is written using the CQL Language [2] with an extension for pattern matching [1]. The inner (nested) query checks for each product if its container is not a freezer or does not exist, and if so retrieves the temperature based on the product's location. The outer query aggregates the retrieved temperatures for the product and checks if it has been exposed to room temperature for 6 hours. The query finally returns all the temperature readings in the 6 hour period and the tag id of the object.

```
Q1:Select tag_id, A[].temp
   From ( Select Rstream(R.tag_id, R.loc, T.temp)
          From    Products [Now] as R, Temperature
                  [Partition By sensor Rows 1] as T
          Where   (!(R.container IsA 'freezer')
                    or R.container = NULL) and
                  R.loc = T.loc and T.temp > 0 °C
        ) As Global Stream S
        [ Pattern SEQ(A+)
          Where  A[i].tag_id = A[1].tag_id and
                 A[A.len].time > A[1].time + 6 hrs
        ]
```

## 3. INFERENCE ALGORITHM

In this section, we present our inference module that translates raw noisy RFID readings, ($time$, $tag\_id$, $reader\_id$), into high-level events with rich attributes ($time$, $tag\_id$, $location$, $container$) and optionally other attributes about object properties from the manufacturer. Our solution to this problem makes use of techniques from probabilistic reasoning, statistics, and machine learning.

---

[1]This technology trend motivated us to minimize the computation state associated with a tag, as discussed in Section 4, so it can be held in a tag's local memory to enable *querying anytime anywhere* in the future.

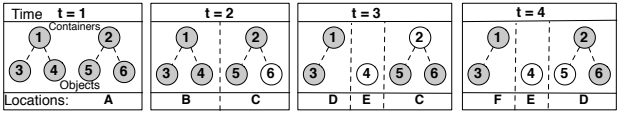

**Figure 1: Example of noisy RFID readings and containment changes**

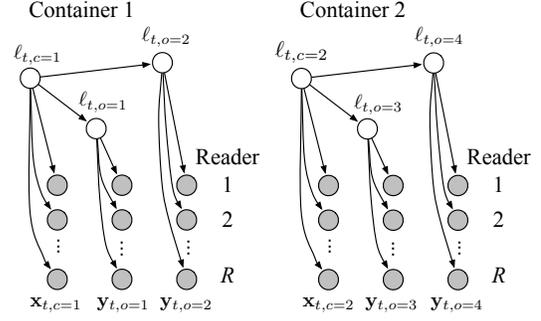

**Figure 2: Graphical model of locations and RFID readings.**

Intuitively, the idea is that whenever an object is read, its container is likely to be read as well. Over time, we can use the *co-location history* of containers and objects to derive the containment relationships. To develop this intuition into a robust system, however, several design considerations must be addressed to effectively handle the noisy and incomplete input. To explain these considerations, we use the example in Figure 1: Each node represents a tag, and each edge a containment relation. The shaded nodes represent tags that were read (by the reader specified in the bottom row), and the unshaded nodes are tags that were missed by all readers.

A main design consideration is how to handle missed readings. If some objects are not observed, it is difficult to accurately determine their locations, which makes it also difficult to tell when objects are co-located. If the containment relations were known for certain, then a powerful way to determine object locations would be to *smooth over containment relations*, meaning that whenever we read one object in a container, we know that all of the other objects must be in the same place. For example, in Figure 1, at time $t = 3$, we miss reading container 2, but we do read object 5. If we knew that container 2 contained object 5, then we could correctly infer that container 2 is also present at location C.

Unfortunately, the containment relationships are not known in advance, so instead we use an *iterative approach*. First, we start with the best available information about object locations and have a guess about containment relationships based on co-location. Then we can improve our understanding of object locations via smoothing over containment relationships. For example, in Figure 1, container 2 and object 5 are repeatedly co-located in the raw readings, so we can infer a containment relationship right away. Given the containment relationship, we can infer the location of container 2 at $t = 3$. The resulting better understanding of locations allows us to further improve our understanding of containment relationships. Revisit Figure 1. We did not have strong evidence about the container for object 6, but with the new location information about container 2, we see that it is consistently co-located with object 6.

A second main design consideration is how to detect changes in containment relationships. Consider an object and a container that have been consistently co-located, such as container 1 and object 4 in the first two time steps of Figure 1. If later on ($t = 3$ in the example), we fail to read the object, then following the idea of smoothing over containment, it is reasonable to infer than object 4 is still co-located with container 1. But at some point, if we repeatedly fail to read the object (as at $t = 4$), we may suspect that the object has actually been moved. To distinguish between these two competing explanations—either the object has been removed from the container, or it has not moved but its tag has been missed—we need a way to decide when there is enough recent evidence to conclude that the containment relationship has actually changed. How much evidence is enough should depend on the read rate: if the readers are less accurate, then we ought to demand more evidence.

We resolve all of these difficulties in a principled way using a general methodology based on graphical modeling. We design a graphical model that represents the probabilistic dependencies between the observed RFID readings and the latent object locations and containment relationships. Inference algorithms infer containment and locations in a unified way, naturally smoothing over the dependencies between them. In the following, we propose a graphical model (Section 3.1) and a new algorithm RFINFER (Section 3.2) for inferring containment and location from RFID readings. Containment changes can be further detected using a change point detection algorithm (Section 3.3). Moreover, inference must run at stream speed, which poses a challenge to existing machine learning methods. The techniques we employ include optimizations (Appendix A.3) and history truncation (Section 4).

## 3.1 Graphical Model

In this section we describe a probabilistic model of container locations, object locations, and RFID readings. The model is a probability distribution over random variables that represent both the true state of the world, which we do not observe, and the RFID readings, which we do. For the purposes of describing the model, we assume that we know the containment relationships exactly; in fact, we infer them from RFID data, as explained in Section 3.2.

We discretize both time and space: We divide time into a set of discrete *epochs* of, for example, one second in duration. All RFID readings that occur in the same epoch are treated as simultaneous. As for locations, given the set of tracking and monitoring queries we aim to support, it suffices to localize objects to the nearest reader. Therefore, we model locations as a discrete set $\mathcal{R}$, which is the set of locations of all of the static readers. Finally, we assume that there are $C$ containers, denoted by integers $c \in [1, C]$, and there are $O$ objects, denoted by integers $o \in [1, O]$.

The random variables in the model are as follows. For each epoch $t$, and each container $c$, let $\ell_{tc}$ be the true location of the container. This is a random variable which takes values from the set of locations $\mathcal{R}$. Similarly, let $\ell_{to}$ be the true location of each object $o$. As for the readings, let $x_{trc}$ be a binary random variable that indicates whether the reader at location $r \in \mathcal{R}$ received a reading of the container $c$. Define $y_{tro}$ similarly for each object $o$. To make the notation more compact, let $\ell = \{\ell_{tc} | \forall t, c\} \cup \{\ell_{to} | \forall t, o\}$ be the vector of all the true object and container locations over all time, and similarly define $\mathbf{x} = \{x_{trc} | \forall t, r, c\}$ for the container readings and $\mathbf{y} = \{y_{tro} | \forall t, r, o\}$ for the object readings. The model is a joint distribution $p(\ell, \mathbf{x}, \mathbf{y})$ over all of these random variables.

Our model is depicted graphically in Figure 2 for a single epoch. To describe the model, we explain how to sample from the probability distribution that describes the world, assuming that the world behaves exactly according to our model. At every epoch $t$, first the true location $\ell_{tc}$ is sampled for each container $c$. Because we do not assume any prior knowledge about the layout of the factory, we model this distribution as uniform over the set of all possible locations $\mathcal{R}$. Now there is no need to sample object locations, because each object must be in the same place as its container.

Now we can generate the RFID readings. Each reader has a *read rate*, which we denote $\pi(r, \bar{r})$, which is the chance of the reader at location $r$ reading an object which is actually at location $\bar{r}$. Typically, a reader detects an object if both are at the same location.

However, with a small chance a reader can detect an object that is closer to a nearby reader. In an actual deployment, one can measure the read rates periodically by using reference tags fixed to known locations and listening for these tags' responses to a given number of interrogations [11, 16]. To create readings, each reader independently interrogates the tag on every container and the tag on every object. Formally, each binary observation variable $x_{trc}$ is sampled independently with probability according to the read rate; that is, $x_{trc}$ is true with probability $\pi(r, \ell_{tc})$. We write this probability as

$$p(x_{trc}|\ell_{tc}) = \begin{cases} \pi(r, \ell_{tc}) & \text{if } x_{trc} = 1 \text{ (tag read)} \\ 1 - \pi(r, \ell_{tc}) & \text{if } x_{trc} = 0 \text{ (otherwise),} \end{cases} \quad (1)$$

and similarly for $y_{tro}$.

Putting it together, this defines a *joint probability distribution* as

$$p(\ell, \mathbf{x}, \mathbf{y}) = \prod_{t=1}^{T} \prod_{c=1}^{C} p(\ell_{tc}) \prod_{r \in \mathcal{R}} p(x_{trc}|\ell_{tc}) \prod_{o|(o,c) \in \mathcal{C}} p(y_{tro}|\ell_{to}) \quad (2)$$

It can be seen that this model treats all time steps as independent and all containers as independent. For each epoch and container, it iterates over all readers and considers the probabilities of each reader observing the container as well as its contained objects. Because the model treats all epochs as independent, it does not perform any temporal smoothing over readings; however, it compensates for this by smoothing over containment relations instead. To smooth the readings over time as well would add significant complexity to the model, and significant computational cost to the inference procedure. In Section 5, we verify experimentally that smoothing over containment relations is effective at inferring object locations.

An important quantity is the probability that the model assigns to the observed data, that is, $p(\mathbf{x}, \mathbf{y}) = \sum_{\ell} p(\ell, \mathbf{x}, \mathbf{y})$. This quantity is called the *likelihood* of the data. Note that the likelihood is a function of the containment relationships $\mathcal{C}$. To emphasize this, we define $L(\mathcal{C}) = \log p(\mathbf{x}, \mathbf{y})$. According to our model, this is

$$L(\mathcal{C}) = \sum_{t=1}^{T} \sum_{c=1}^{C} \log \sum_{a \in \mathcal{R}} p(\ell_{tc}) \prod_{r \in \mathcal{R}} p(x_{trc}|\ell_{tc}) \prod_{o|(o,c) \in \mathcal{C}} p(y_{tro}|\ell_{to}) \quad (3)$$

The log likelihood measures how probable the RFID readings are under the current set of containment relationships. It will be an important quantity for inferring the containment relationships.

### 3.2 Inferring Containment Relationships

To infer containment relationships from RFID readings, we use a *maximum likelihood* framework, that is, we determine the containment relationships such that, according to the model, the observed readings are most likely. Formally, this amounts to maximizing the log likelihood $L(\mathcal{C})$ with respect to the set of containment relationships $\mathcal{C}$. In this section, we describe the algorithm that performs this maximization, which we call RFINFER.

The idea is that determining containment relationships would be simple if, besides the RFID data, we also observed the true locations of all containers. However, the true container locations are in fact unknown. To handle this, we develop RFINFER in the EM framework, which offers a general approach for maximizing likelihood functions in the presence of missing data, in our case the container locations. The algorithm alternates between two steps. In the first step, the *expectation step* (or E-step), we infer a distribution over the locations of each container, given some current guess about the containment relations. In the second step, the *maximization step* (or M-step), we choose the best containment relations given our current guess of the container locations. We iterate these two steps until the containment relations do not change.

In the E-step, the distribution that we want to compute is the conditional distribution $p(\ell|\mathbf{x}, \mathbf{y})$ over the location of each container that results from the joint distribution of Eq (3)—this distribution is called the *posterior distribution* of the container location and sometimes denoted as $q_{tc}(\cdot)$ for simplicity. From the definition of conditional probability, it can be shown that

$$p(\ell_{tc}|\mathbf{x}, \mathbf{y}) = S \prod_{r \in \mathcal{R}} p(x_{trc}|\ell_{tc}) \prod_{o|(o,c) \in \mathcal{C}} p(y_{tro}|\ell_{to}), \quad (4)$$

where $S$ is a constant that does not depend on $\ell_{tc}$.

In the M-step, we update the current estimates of containment relationships based on the current belief about locations. We do so by defining a score $w_{co}$ to measure the *strength of co-location* between object $o$ and container $c$:

$$w_{co} = \sum_{t=1}^{T} \sum_{a \in \mathcal{R}} p(\ell_{tc} = a|\mathbf{x}, \mathbf{y}) \sum_{r \in \mathcal{R}} \log p(y_{tro}|\ell_{to} = a). \quad (5)$$

This score measures how likely are the readings of object $o$ if it were always co-located with container $c$. To estimate the container for $o$, we simply pick the best container $C(o) = \arg\max w_{co}$.

Note that RFINFER also computes location information. When the algorithm has converged, the final values of $p(\ell_{tc}|\mathbf{x}, \mathbf{y})$ are our best estimates of the location of each container at each time step and the locations of objects believed to be in the container.

Finally, the following theorem states that our algorithm is guaranteed to converge to an optimum of the likelihood:

**Theorem 1.** *The* RFINFER *algorithm converges, and the resulting values* $\mathcal{C}^*$ *are a local maximum of the likelihood defined in Eq* (3).

The proof is given in the Appendix A. The key step is to show that our simple, custom M-step indeed maximizes the likelihood.

**Complexity, Optimizations, and Extensions.** We refer the reader to the Appendix A for the complexity analysis, implementation, optimizations, and extensions of our algorithm. After a series of optimizations, our algorithm achieves a linear complexity, $O(C + O)$, in each iteration, and usually converges in just a few iterations.

### 3.3 Change Point Detection

In this section, we describe how we infer changes in containment relationships. This type of problem, called *change-point detection*, is the subject of a large literature in statistics (see [3] for an overview). A *change point* is a time $t$ at which the containment relationships change, that is, some object has either changed containers or been removed altogether. Finding change points is challenging because of the noise in RFID readings. For example, in Figure 1 at $t = 4$, it may be unclear if object 4 has actually been removed from container 1, or it has simply been "unlucky" enough to be missed twice in a row. To distinguish these two possibilities, we need a way to quantify the unluckiness of a set of readings.

We propose a statistical approach based on hypothesis testing. Suppose that we have received readings from epochs $[0, T]$. Then we define a *null hypothesis*, which is that the containment relationships have not changed at all during epochs $[0, T]$. Then, if under the null hypothesis, it turns out that the observed RFID readings are highly unlikely, we reject the null hypothesis, concluding that a change point has in fact occurred. To measure whether the observed readings are unlikely, we again use the likelihood Eq (3). Consider a single object $o$. Let $\mathcal{C}_{0:T}$ be the maximum likelihood containment relations based on the full data, so that $L(\mathcal{C}_{0:T})$ is the best possible likelihood if there is no change point. Alternatively, suppose there is a change point at some time $t'$. Then let $\mathcal{C}_{0:t'}$ and $\mathcal{C}_{t':T}$ be the best containment relations that allow object $o$ to change locations at time $t'$. Maximizing over possible change points, the best possible likelihood if there is any change point for $o$ is $\max_{t'} L(\mathcal{C}_{0:t'}) + L(\mathcal{C}_{t':T})$. We perform change point detection using the difference of these two log likelihoods, that is,

$$\Delta_o(T) = L(\mathcal{C}_{0:T}) - \max_{t' \in [0,T]} [L(\mathcal{C}_{0:t'}) + L(\mathcal{C}_{t':T})] \qquad (6)$$

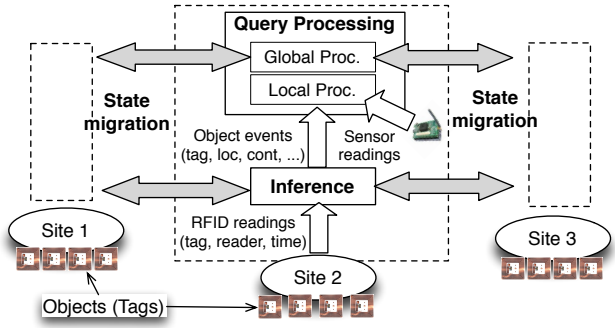

**Figure 3: A distributed RFID data management system.**

Essentially, this measures how much better we can explain the data if we use two different sets of containment relationships instead of one. This is a type of *generalized likelihood ratio* statistic, which is a fundamental tool in statistics. The change point detection procedure will signal that there has been a change point whenever the value of $\Delta_o(T)$ is greater than a threshold $\delta$.

Intuitively, to choose the threshold we would like to know what values of $\Delta_o(T)$ would be typical if there were no change point. Fortunately, we can obtain as much of this data as we want, simply by sampling hypothetical observation sequences from the model, exactly as described in Section 3.1. Since none of the hypothetical sequences actually contain a change point, if our procedure signals a change point on one of them, it must be a false positive. In practice, all of the hypothetical $\Delta_o(T)$ values are quite small, so we choose $\delta$ to be their maximum. Furthermore, all of this computation can be done in advance before any RFID data is observed. The details of the change point detection procedure are given in Appendix A.2.

## 4. DISTRIBUTED PROCESSING

As object tracking and monitoring systems grow into many geographically separate sites and millions of objects, the sheer volume of data poses a scalability challenge. A centralized approach, like centralized warehousing, requires all the data to be transferred to a single location for processing. This approach incurs both delay of answering queries and high communication costs.

In this work, we propose a distributed approach natural for object tracking and monitoring, which performs *"querying where an object (and data) is located"*. The architecture of such a distributed system is illustrated in Figure 3. As can be seen, each site performs inference and query processing on local RFID streams as objects are observed. Inference runs on raw RFID streams and produces an object event stream describing the location and container of each object. Query processing runs continuously on the object event stream and other sensor streams to return all answers. Inference and query processing, however, often require information from the previous sites that an object has passed. To solve this problem, we perform *state migration*, which transfers the state of inference and query processing for an object when it moves across sites.

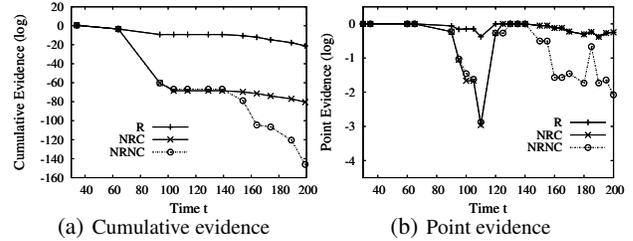

(a) Cumulative evidence    (b) Point evidence

**Figure 4: Evidence of co-location of three candidate containers.**

State migration can be realized in several ways: ($i$) When an object is scanned at the exit of a site, if domain knowledge about its next location is available, its inference and query processing state can be transferred directly to that site. ($ii$) Alternatively, when an object reaches a new site, the server there can locate the object's previous place using the Object Naming Service (ONS) and retrieves its state from that place. ($iii$) Finally, it is desirable to write the object's state to the local storage of the RFID tag (once the technology of writable tags matures for large deployments), while leaving a copy of the state at the current site as backup. This method will enable *querying instantly when a tag is in sight*, with minimum delay of answering queries and minimum communication costs.

To reduce communication costs or cope with limited local tag storage, it is important to minimize inference and query processing state while ensuring accuracy of query answers. We address this issue in both inference and query processing as described below.

### 4.1 State Migration for Inference

Our inference algorithm presented in the previous section requires the entire *history* of readings associated with each object produced from all the sites that this object has passed. When an object leaves one site for another, the history of this object and the history of all of its possible containers, collectively called the *inference state* of the object, need to be transferred to the new location for subsequent inference. Evidently, transferring the complete history of objects and containers would incur both a high communication cost across sites and a high processing cost at the new location. Below, we describe two techniques to address these problems.

**Truncating History**. The goal of *history truncation* is to sift out the observations that are most informative about true containment relationships from history, and retain only those for future processing. This can be accomplished by monitoring the strength of co-location computed in our containment inference algorithm RFINFER. Recall from Eq (5) for the M-step of the RFINFER algorithm, the co-location strength $w_{co}$ for each object $o$ and container $c$ is a sum over all time steps of a quantity which we call the *point evidence* of co-location. We denote this quantity by:

$$e_{co}(t) = \sum_{a \in \mathcal{R}} q_{tc}(a) \sum_{r \in \mathcal{R}} \log p(y_{tro}|\ell_{to} = a). \qquad (7)$$

Then the *cumulative evidence* of co-location can be computed as $E_{co}(t) = \sum_{t'=1}^{t} e_{co}(t')$.

To see how these quantities are used, suppose that in a warehouse an object started at the entry door at time 0, was scanned on the conveyor belt around time 100, and then placed on a shelf at time 150. Consider three candidate containers that were co-located with this object at the entry door: the real container (denoted by $R$) always traveled with the object; a second container ($NRC$) was co-located at the door and at the shelf, but not at the belt; a third container ($NRNC$) was not co-located after the door. Figure 4(a) shows the cumulative evidence of co-location of three candidate containers with the object. Around time 100, the belt reader scanned the real container alone with the object, causing the cumulative evidence of the other two containers to drop fast. This is exactly the informative region we want to find in history truncation. The information afterwards is less useful, because the false container $NRC$ is co-located with the object again on the shelf, while the false container $NRNC$ was already eliminated from contention by the belt reader.

Our history truncation algorithm aims to find a time period, called the *critical region*, whose observations are most informative for determining containment. While our intuition was explained using the cumulative evidence of co-location, our algorithm actually uses the point evidence of co-location, as shown in Figure 4(b) (in log space). During the critical region around time 100, the real container has much higher point evidence than the two false containers; this is not true either before or after the region.

After containment inference completes, our history truncation algorithm runs as follows: It searches through time by applying a small sliding window $[t-w, t]$. Given the current window, for each object $o$, it computes the sum of point evidence $\sum_{t'=t-w}^{t} e_{co}(t')$ for each possible container of $o$. If the difference in sum between the best container and the second best is large enough (using a heuristic-based threshold), the current window is considered a critical region $CR$ of the object, overwriting the previous $CR$ if existent. When the search reaches the end, the most recent $CR$ is the final critical region of the object. Readings of the object and its possible containers outside the critical region will be all ignored.

After running the algorithm, we have compressed the entire history from $[0, T]$ to a small region $CR$. When the new readings arrive in the time period $[T, T']$, rather than running inference over the entire period $[0, T']$, we run inference only over the data in the the critical region $CR$ and in recent history denoted by $\bar{H}$. If containment is stable, it suffices to have $\bar{H} = [T, T']$, i.e., including all new readings obtained since last inference. To support change point detection, however, we may need a somewhat larger recent history $\bar{H}$. According to Eq (6), the change point can be any point in the entire history. In practice, it is more likely to be in the recent history since it was not detected last time. However, it may be imprudent to restrict the change point only to the most recent period $[T, T']$ because a change point before the time $T$ might not get sufficient evidence in the previous change point detection. Our experimental results in Section 5.1 show that the sufficient size of $\bar{H}$ is within a factor of 2 of $T'$-$T$. As time elapses, the recent history $\bar{H}$ moves forwards and we can truncate the readings falling behind $\bar{H}$ by applying the critical region algorithm again.

**Collapsing Inference State.** When an object leaves a site for the next, the *inference state* for the object includes the readings of the object and the readings of its candidate containers in both the critical region $CR$ and the recent history $\bar{H}$. One solution is simply shipping the inference state to the next site to seed inference there. However, the inference state for an object may not be small since each object can have dozens of candidate containers, and each container or object can have hundreds of readings in $CR$ and $\bar{H}$.

In our work we employ a technique to collapse the inference state to a single number for each container-object pair, i.e., the co-location weight $w_{co}$, hence avoiding the overhead of transferring readings entirely. This dramatically reduces the inference state transferred between sites. Then the inference algorithm at a new location simply adds the old transferred weights to the new weights that are computed from the readings at the new site. This technique, however, can affect accuracy: if later evidence shows that the containment inference results from the old location were incorrect, we can no longer revise the old estimates as the corresponding readings have been discarded. Even in this case, however, inference in the new place still has a chance to correct the old estimates because readings obtained there will eventually overrule the old weights.

## 4.2 State Migration for Querying

Given an event stream with object location and containment information, the query processor processes this stream and other sensor streams to answer monitoring queries. Our discussion below assumes CQL-based relational stream processing [2] extended with the pattern matching functionality [1].

Under our approach "querying where an object is located", a monitoring query is registered with every site. It is split into local processing and global processing parts based on the labels of input streams specified in the query. For each query block, if any of the input streams is labeled as "global", then this block uses global processing across sites; otherwise, it is processed only on the local streams. While local processing can be performed by existing stream systems [1, 2], global query processing requires additional mechanisms. First, global query processing needs to maintain *computation state* for each object. Since all stream systems maintain computation state (a.k.a. synopsis [2]) and update it with each arriving tuple, our work further partitions the state according to individual objects. Then as an object leaves one site for another, we perform *state migration* using one of the three strategies mentioned at the beginning of the section. See Appendix B for illustration of the above approach using Query 1 in Section 2.

A main issue in state migration is that the total amount of state to be transferred can be enormous given a large number of objects. To reduce communication costs, we exploit *stable containment* to share query states across objects. At the exit point of a storage area, we consider the objects in each container, e.g., frozen food products considered in Query 1. These objects have the same container and location at present (but possibly different histories). The query states for these objects are likely to have commonalities. Hence, we propose a centroid-based sharing technique that finds the most representative query state and compresses other similar query states by storing only the differences. Details are available in Appendix B.

## 5. PERFORMANCE EVALUATION

We have implemented a prototype of our inference approach, (including the optimizations in Appendix A), connected it to a stream query processor [1], and extended both to distributed processing. We evaluate our system using both synthetic traces emulating RFID-based supply chains and real traces from a laboratory setup.

## 5.1 Single-Site Inference

We first evaluate our inference algorithm on synthetic RFID streams from a single warehouse. The detailed experimental setup, performance metrics, and additional results are given in the Appendix C. By default, we run inference every 300 seconds.

**Inference with stable containment.** We evaluate our inference algorithm first using traces with stable containment. To deal with traces of various lengths, we consider the Critical Region (CR) method that we proposed for history truncation in distributed processing (Section 4.1) as an optimization also for traces produced at a single warehouse. This method results in the use of the critical region and a short recent history $\bar{H}$ (by default, the most recent 600 seconds) for inference. For comparison, we also include a simple window-based truncation method that keeps the most recent $W$ readings for inference ($W$=1200 seconds here).

We first test the sensitivity of these methods to the read rate $RR$. As Figure 5(a) shows, while all three methods offer high accuracy for location inference (all three lines for location inference are very close, so we show only the line for Location(CR) for readability), they differ widely for containment inference: The window method has the worse accuracy because when the useful observations, such

331

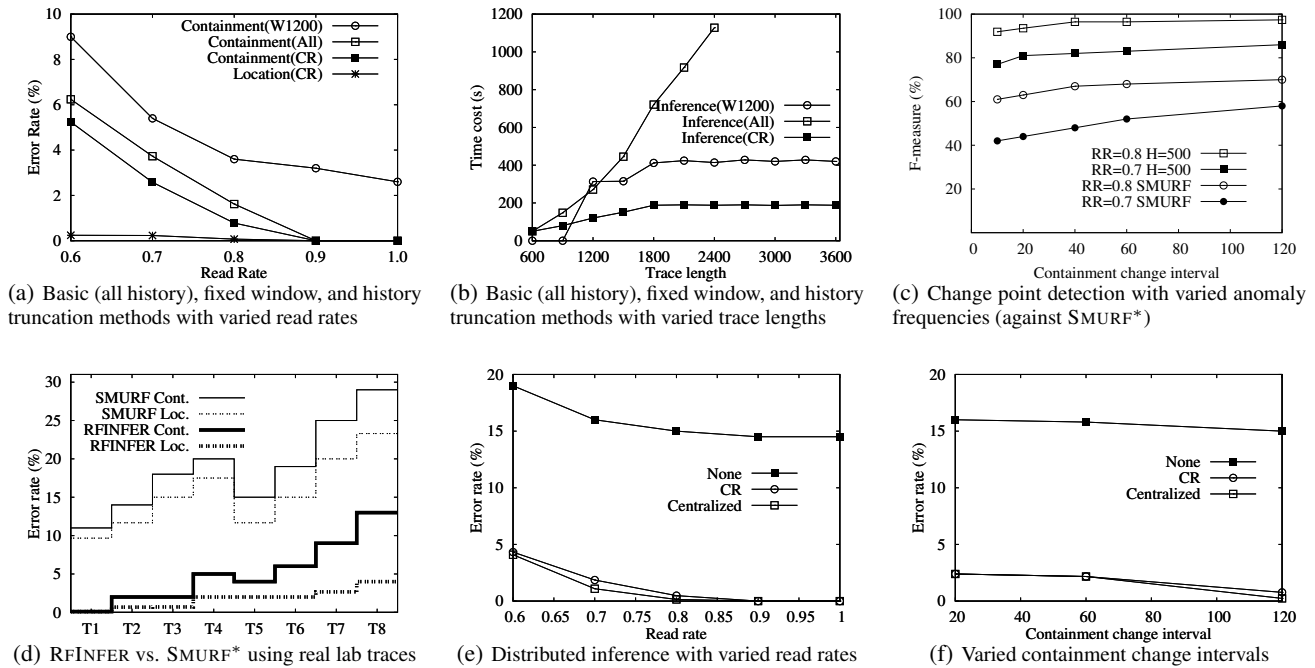

**Figure 5:** Experimental results for single-site inference (*a-c*), our lab warehouse deployment (*d*), and distributed inference (*e* and *f*).

as the belt readings, fall outside the window, the inference algorithm can no longer use them to infer containment. Using the entire history or the CR method gives better accuracy as expected. Interestingly, while the CR method was initially proposed for improving performance, it also improves over the basic algorithm in accuracy due to the removal of noise (e.g., co-location of a false container and an object on a shelf) from inference. Moreover, its sensitivity to the read rate is comparable to that using the full history, which is the best that one can expect.

We next vary the trace length from 600 to 3600 seconds and compute the total inference time when using the entire history, the window, and the CR methods in Figure 5(b). Here we see that using the entire history severely penalizes the performance, the window based truncation stays in the middle, and the CR method performs the best with its running time insensitive to the trace length.

**Containment change detection.** We next employ the change point detection algorithm from Section 3.3 to detect containment changes. We use a recent history size of $\bar{H}$ (600 seconds by default) in addition to the detected critical region for inference with change point detection. To generate events of interest, we inject anomalies that randomly choose an item and move it to a different case in the warehouse. The frequency of such anomalies is every 20 seconds by default but also varied over a wide range. Each run simulates a warehouse with 32,000 items in steady state over 4 hours.

*Choice of threshold.* We first examine the effect of the threshold $\delta$ for change point detection. We consider fixed values in a range as well as our offline method as described in Section 3.3. Due to space constraints, the details of this study are left to Appendix C.4 (see Table 3). In summary, our chosen threshold always approximates the optimal value within 2% across all read rates.

*Tradeoff between accuracy and efficiency.* We further study the tradeoff between accuracy and efficiency. The change point detection algorithm requires a recent history (whose size is $\bar{H}$), besides the critical region in the past, to detect containment changes. Results of our study show that a longer recent history helps improve accuracy especially when read rates are low, while it may increase inference cost. Overall, our algorithm can achieve 85% accuracy given read rates $\geq 0.7$ while keeping up with stream speed (by using a relatively small $\bar{H}$). The details are shown in Appendix C.4.

*Frequency of unexpected containment changes.* We next test the sensitivity of our algorithm to the frequency of unexpected containment changes, i.e., without using special readers to scan containers separately. We varied the interval between two containment changes, from 10 to 120 seconds. For comparison, we include an alternative method, called SMURF*, that extends the state-of-the-art SMURF[11] method for RFID data cleaning with heuristics for containment inference (see Appendix C.3 for details). For our algorithm, we chose the $\bar{H}$ size to keep up with stream speed based on Table 4 in Appendix C.4, i.e., $\bar{H}=500$ for both $RR=0.7$ and $RR=0.8$. As Figure 5(c) shows, our algorithm is much more accurate than SMURF* and is not very sensitive to the containment change interval. SMURF* is much worse because it lacks a principled approach to exploiting the iterative feedback between location and containment estimates.

## 5.2 Evaluation of Lab RFID Deployment

To evaluate our system in real-world settings, we developed an RFID lab with 7 readers and 20 cases containing 5 items each to simulate a small warehouse. We created 8 traces, labeled $T_1, \ldots, T_8$, with different characteristics regarding the environmental noise and overlap among readers (for details see Appendix C.2). We ran inference every 5 minutes using a 10-minute history (or all the data available if the history is less than 10 minutes). For comparison, we include the SMURF* algorithm described in Appendix C.3.

Figure 5(d) shows the inference error rates for RFINFER and SMURF*. As can be seen, RFINFER is much more accurate than SMURF* across all traces although they both use intuitions such as smoothing and co-location. Again, this is because RFINFER uses *smoothing over containment relations* and a principled approach for the iterative feedback between location and containment estimates. This is shown to be more effective than smoothing over time for individual objects and then combining such location evidence in a heuristic way to infer containment as in SMURF*. For RFINFER, the location error rates are low across all traces. In the absence of containment changes, the containment error rates are within 5% in traces $T_1$ to $T_4$ despite the heterogeneous read rates, added environmental noise, and significant overlap between readers. Containment changes cause containment error rates to rise, especially given lower read rates or higher overlap rates, but with a maximum of 13% with all the noise factors combined in $T_8$.

## 5.3 Distributed Inference

**Accuracy and Communication Cost**. We next compare centralized and distributed approaches to inference by simulating 10 warehouses for 4 hours. Each warehouse has 32,000 items in steady state, totally 0.32 million items. Our system runs inference at stream speed for each warehouse. Figure 5(e) shows the error rates for varied read rates. The naive no state-transfer method (labeled "None") has a high error rate, while our critical region (CR) method perform close to the centralized method. Figure 5(f) shows similar results when the containment change frequency varies. The communication costs (shown in Appendix C.5) show that our CR methods offer 3 orders of magnitude reduction in communication cost over a centralized approach while approximating its accuracy.

**Scalability**. We further test the scalability of our inference system by using larger numbers of objects in simulation. Our inference system can scale to 150,000 items per warehouse while keeping up with stream speed, totaling 1.5 million objects over 10 warehouses. The above reported results on accuracy and communication costs stay true. One way to support more objects is to use mobile readers for scanning objects on shelves (which is a more cost-effective deployment than static readers). In another simulation, we use a mobile reader to scan each isle of 90 shelves. The mobile reader reads every second and spends 10 seconds scanning each shelf. Given such reduced shelf readings, our inference system can scale to 1.21 million items per warehouse while running at stream speed, totaling 12.1 million objects over 10 warehouses.

## 5.4 Distributed Inference and Querying

We finally extend our distributed inference experiment with query processing. We report results using two representative queries: Q1 from Section 2, and Q2 that reports the frozen food that has been exposed to temperature over 10 degrees for 10 hours. The table below reports the F-measure of query results and the total size of query state with and without the containment-based sharing method (Section 4.2). We see that the overall accuracy of query results is high ($> 89\%$). Also, state sharing yields up to 10x reduction in query state size. Finally, the accuracy and query state reduction ratio of Q1 are lower than those of Q2. This is because Q1 combines inferred location and containment, but Q2 only uses the inferred location which is more accurate than the inferred containment.

|    |                      | RR=0.6 | RR=0.7 | RR=0.8 | RR=0.9 |
|----|----------------------|--------|--------|--------|--------|
|    | F-m.(%)              | 89.2   | 94     | 95.1   | 96     |
| Q1 | State w/o share(bytes)| 65,500 | 66,000 | 67,037 | 67,000 |
|    | State w. share(bytes)| 6,986  | 5,737  | 5,589  | 5,156  |
|    | F-m.(%)              | 93.5   | 96.1   | 97.3   | 97.5   |
| Q2 | State w/o share(bytes)| 80,248 | 85,510 | 87,029 | 87,000 |
|    | State w. share(bytes)| 7,296  | 6,108  | 5,341  | 5,273  |

## 6. RELATED WORK

*RFID stream processing*. Recent research has addressed RFID data cleaning [8] and location inference for static readers [11, 14, 4] and mobile readers [16]. However, containment inference is more challenging since inter-object relationships cannot be directly observed. Our work is the first to employ *smoothing over object relations* in RFID inference, with demonstrated performance. Our work further supports distributed inference and querying.

*RFID databases*. Existing work has addressed RFID data archival [17], event specification and extraction [18], integrating data cleansing with query processing [13], and exploiting known constraints to derive high-level information [19]. Our system addresses a different problem: it processes raw data streams to infer object location and containment, thereby enabling stream query processing, and scales inference and query processing to distributed environments.

*Inference in sensor networks*. Various techniques [12, 6, 15, 10] have been used to infer true values of temperature, light, object positions, etc., that a sensor network is deployed to measure. Our inference problem differs because the inter-object relationships, such as containment, cannot be directly measured, hence requiring different statistical models and inference techniques. We further address distributed inference and query processing for scalability.

## 7. CONCLUSIONS

In this paper, we presented the design of a scalable, distributed stream processing system for RFID tracking and monitoring. Our technical contributions include ($i$) novel inference techniques that provide accurate estimates of object locations and containment relationships in noisy, dynamic environments, and ($ii$) distributed inference and query processing techniques that minimize the computation state transferred across warehouses while approximating the accuracy of centralized processing. Our experimental results demonstrated the accuracy, efficiency, and scalability of our techniques. In future work, we plan to extend our work to include probabilistic query processing, exploit on-board tag memory to hold object state and enable anytime anywhere querying, and explore smoothing over object relations for other data cleaning problems.

## 8. REFERENCES

[1] J. Agrawal, Y. Diao, et al. Efficient pattern matching over event streams. In *SIGMOD*, 147–160, 2008.
[2] A. Arasu, et al. The CQL continuous query language: semantic foundations and query execution.. VLDB J. , 15(2): 121-142, 2006.
[3] M. Basseville and I. V. Nikiforov. *Detection of Abrupt Changes: Theory and Application*. Prentice-Hall, 1993.
[4] H. Chen, W.-S. Ku, et al. Leveraging spatio-temporal redundancy for RFID data cleansing. In *SIGMOD '10*, 51–62, 2010.
[5] N. N. Dalvi and D. Suciu. Efficient query evaluation on probabilistic databases. *VLDB J.*, 16(4):523–544, 2007.
[6] M. Cetin, L. Chen, et al. Distributed fusion in sensor networks. *IEEE Signal Processing Mag.*, 23:42–55, 2006.
[7] K. Finkenzeller. *RFID handbook: radio frequency identification fundamentals and applications*. John Wiley and Sons, 1999.
[8] M. J. Franklin, S. R. Jeffery, et al. Design considerations for high fan-in systems: The HiFi approach. In *CIDR*, 290–304, 2005.
[9] S. Garfinkel and B. Rosenberg, editors. *RFID: Applications, Security, and Privacy*. Addison-Wesley, 2005.
[10] A. Ihler, J. Fisher, et al. Nonparametric belief propagation for self-calibration in sensor networks. In *IPSN*, 225–233, 2004.
[11] S. R. Jeffery, et al. An adaptive RFID middleware for supporting metaphysical data independence. *VLDB Journal*, 17(2):265–289, 2007.
[12] M. Paskin, C. Guestrin, et al. A robust architecture for distributed inference in sensor networks. In *IPSN*, 55–62, 2005.
[13] J. Rao, S. Doraiswamy, et al. A deferred cleansing method for RFID data analytics. In *VLDB*, 175–186, 2006.
[14] C. Ré, J. Letchner, et al. Event queries on correlated probabilistic streams. In *SIGMOD*, 715–728, 2008.
[15] J. Schiff, D. Antonelli, et al. Robust message-passing for statistical inference in sensor networks. In *IPSN*, 109–118, 2007.
[16] T. Tran, C. Sutton, et al. Probabilistic inference over RFID streams in mobile environments. In *ICDE*, 1096-1107, 2009.
[17] F. Wang and P. Liu. Temporal management of RFID data. In *VLDB*, 1128–1139, 2005.
[18] E. Welbourne, et al. Cascadia: a system for specifying, detecting, and managing RFID events. In *MobiSys*, 281–294, 2008.
[19] J. Xie, J. Yang, et al. A sampling-based approach to information recovery. In *ICDE*, 476–485, 2008.

| | |
|---|---|
| $R$ | Number of reader locations |
| $C$ | Number of containers |
| $O$ | Number of objects |
| $o$ | Index of a single object; $o \in [1, O]$ |
| $c$ | Index of a single container; $c \in [1, C]$ |
| $t$ | Index of time epoch (e.g., 1 second long) |
| $\mathcal{R}$ | Set of possible reader locations |
| $\ell_{tc}$ | True location of container $c$ at time $t$. |
| $\ell_{to}$ | True location of object $o$ at time $t$ |
| $\pi(r, \bar{r})$ | Read rate. Probability that reader at location $r \in \mathcal{R}$ detects an object at location $\bar{r} \in \mathcal{R}$ |
| $y_{tro}$ | Binary variable indicating whether object $o$ was read by reader at location $r$ at time $t$ |
| $x_{trc}$ | Binary variable indicating whether container $c$ was read by reader at location $r$ at time $t$ |
| $\mathbf{x}$ | Binary vector of all container readings |
| $\mathbf{y}$ | Binary vector of all object readings |
| $w_{co}$ | Strength of co-location between container $c$ and object $o$ |
| $\mathcal{C}$ | Containment relations; set of pairs (`object_id`, `container_id`) |
| $L(\mathcal{C})$ | Likelihood of the observed readings, given containment relations $\mathcal{C}$ |
| $\Delta_o(T)$ | Change-point statistic for epoch $T$ |

Table 1: Notation used in this paper

# APPENDIX

The notation used in this paper is summarized in Table 1.

## A. ENHANCEMENTS OF RFINFER

Below we present additional details about our inference algorithm, its implementation and optimizations, and two extensions.

### A.1 Pseudocode and Proof of RFINFER

The pseudocode for RFINFER is shown in Algorithm 1. We next prove Theorem 1 about the optimality of the RFINFER algorithm.

*Proof.* We show that RFINFER (Algorithm 1) is guaranteed to converge to a local maximum of the likelihood $L(\mathcal{C})$ in (3). Following the EM theory, we can interpret both the E-step and the M-step as maximizing a lower bound on the likelihood, which is

$$L(\mathcal{C}) \geq \sum_{t=1}^{T} \sum_{c=1}^{C} \sum_{a \in \mathcal{R}} q_{tc}(a) \log \frac{p(\ell_{tc} = a, \mathbf{x}, \mathbf{y})}{q_{tc}(a)} = \mathcal{O}(\mathcal{C})$$

The fact that this is a lower bound can be proven by Jensen's inequality. The E-step maximizes this bound with respect to $q_{tc}$, and the M-step with respect to $\mathcal{C}$. In RFINFER, the E-step is identical to the standard E-step of EM, but we use a custom M-step that is specific to our model. So it suffices to prove that the M-step in RFINFER indeed maximizes $\mathcal{O}(\mathcal{C})$.

When maximizing with respect to $\mathcal{C}$, we can ignore terms that do not depend on $\mathcal{C}$. Expanding $\mathcal{O}(\mathcal{C})$ using Eq (3) and removing irrelevant terms yields

$$\max_{\mathcal{C}} \mathcal{O}(\mathcal{C}) = \max_{\mathcal{C}} \sum_{t=1}^{T} \sum_{c=1}^{C} \sum_{a \in \mathcal{R}} q_{tc}(a) \sum_{o|(o,c) \in \mathcal{C}} \log p(\mathbf{y}_{tc}|\ell_{to} = a)$$

$$= \max_{\{c(o), \forall o\}} \sum_{o=1}^{O} w_{c(o), o},$$

where $c(o)$ denotes the container of object $o$, and $\mathbf{y}_{tc} = \{y_{trc}|\forall r\}$. In this last equation, notice that each containment decision $c(o)$ that we are maximizing over appears in only one term of the summation. This means that we can find the global maximum by maximizing each term independently, i.e., $\max_{\mathcal{C}} \mathcal{O}(\mathcal{C}) = \sum_o \max_{c'} w_{c', o}$. This is exactly what is computed in lines 12–20 of RFINFER. □

**Algorithm 1** Pseudocode of RFINFER for inferring containment

while not converged do
　// E step: compute new q
　for $t = 0$ to $T$ do // For each epoch
　　for $c = 1$ to $C$ do // For each container
5　　　for all $a \in \mathcal{R}$ do // For all possible locations
$$q_{tc}(a) \leftarrow \prod_{r \in \mathcal{R}} p(x_{trc}|\ell_{tc} = a) \prod_{o|(o,c) \in \mathcal{C}} p(y_{tro}|\ell_{to} = a)$$
　　　// Now $q_{tc}(a) = S^{-1} p(\ell_{tc} = a|\mathbf{x}, \mathbf{y})$
　　　$S \leftarrow \sum_{a \in \mathcal{R}} q_{tc}(a)$
　　　for all $a \in \mathcal{R}$ do // For all possible locations
10　　　　$q_{tc}(a) \leftarrow q_{tc}(a)/S$
　　　// Now $q_{tc}(a) = p(\ell_{tc} = a|\mathbf{x}, \mathbf{y})$
　// M step: compute new $w$
　for $o = 1$ to $O$ do // For each object
　　for $c = 1$ to $C$ do // For each container
15　　　$w_{co} \leftarrow \sum_{t=1}^{T} \sum_{a \in \mathcal{R}} q_{tc}(a) \sum_{r \in \mathcal{R}} \log p(y_{tro}|\ell_{to} = a)$
　// M step: compute new containment set
　$\mathcal{C} \leftarrow \emptyset$
　for $o = 1$ to $O$ do // For each object
　　$c^* \leftarrow \arg\max_{c \in [1, C]} w_{co}$
20　　$\mathcal{C} \leftarrow \mathcal{C} \cup \{(o, c^*)\}$

**Computation Complexity.** Each iteration of RFINFER requires $O(TCOR^2)$ time, where by *iteration* we mean a single execution of lines 2–20. This is due to two reasons. First, the computation of $q_{tc}(a)$ in line 6 requires $O(OR)$ time, and is executed $O(TCR)$ times by the outer loops. Second, the computation of $w_{co}$ in line 15 requires $O(TR^2)$ time, and is executed $O(CO)$ times by its outer loops. This is the running time of a naive implementation of the algorithm; in Appendix A.3, we describe several optimizations that improve the performance significantly. Also, note that this is the computational complexity per iteration. In general, it is difficult to characterize the number of iterations required for EM to converge, because this depends strongly on characteristics of the unknown true distribution. However, we observe empirically that our inference algorithm usually converges in just a few iterations.

### A.2 Details of Change Point Detection

The change point detection procedure works as follows: First, before any data arrives, choose the threshold $\delta$ as described in Section 3.3. Then, the change point detection is run after each time the RFINFER algorithm runs, which also provides the null hypothesis. For each object $o$, we compute

$$\Delta_o(T) = L(\mathcal{C}_{0:T}) - \max_{t' \in [0, T]} \left[ L(\mathcal{C}_{0:t'}) + L(\mathcal{C}_{t':T}) \right]$$

If $\Delta_o(T) < \delta$, then there is no change point for $o$. Otherwise, if $\Delta_o(T) \geq \delta$, then we flag a change point at the time $t'$ that achieved the maximum in Eq (6). Moreover, we disregard the data from $0 \ldots t'$ in all subsequent calls to the change point algorithm, so we do not flag the same change point more than once.

In implementation, this procedure incurs little extra cost beyond the computation in the RFINFER algorithm. Recall that in the M-step of the RFINFER algorithm, the co-location strength $w_{co}$ for each object $o$ and container $c$ is a sum over all time steps of a quantity that we call the point evidence of co-location, denoted

$$e_{co}(t) = \sum_{a \in \mathcal{R}} q_{tc}(a) \sum_{r \in \mathcal{R}} \log p(y_{tro}|\ell_{to} = a).$$

The cumulative evidence of co-location is $E_{co}(t) = \sum_{t'=1}^{t} e_{co}(t')$. The point evidence of object $o$ and container $c$ at each time $t$ is memorized so it can be re-used to calculate $L(\mathcal{C}_{0:t'})$ and $L(\mathcal{C}_{t':T})$: $L(\mathcal{C}_{0:t'}) = E_{co}(t')$ and $L(\mathcal{C}_{t':T}) = \sum_{t=t'}^{T} e_{co}(t)$. If the change point is detected at time $t'$, we use the $w'_{co} = \sum_{t=t'}^{T} e_{co}(t)$ as the new strength of co-location to get the new container for object $o$. All the computation above is simply the sum of the point evidence memorized from the computation of the RFINFER algorithm. Hence change point detection incurs little extra overhead.

### A.3 Implementation and Optimizations

In this section, we sketch the main data structures and optimizations that we use to implement the RFINFER algorithm including the change point detection extension.

**Data structures**. We use a series of tables: (1) The read rate table (size: $R \times R$) stores the read rates $\pi(r, \bar{r})$. (2) Two history tables: one for container readings **x** (size: $T \times R \times C$), and one for object readings **y** (size: $T \times R \times O$) (3) The posterior probability table ($T \times C \times R$) stores the posterior distribution $q_{tc}(a)$ over container locations. (4) The weight table ($C \times O$) stores the co-location strengths $w_{co}$. (5) Finally, the containment table (vector of length $O$) stores the container inferred for each object. Although the history and posterior probability tables grow with time, in Section 4 we describe a history truncation method that reduces the memory requirement without sacrificing accuracy. Finally, many of these tables, especially the history tables, are sparse, i.e., most cells are 0, and so can be easily compressed to save memory.

**Optimizations**. We further employ several optimizations to improve inference efficiency. Recall from Appendix A.1 that both the E-step and M-step of our algorithm have the complexity $O(TCOR^2)$. The E-step can be easily improved as each object is typically read in only a small number of locations and each container contains a small number of items. Since both of those are bounded independently of $R$ and $O$, the E-step can be improved by a factor of $OR^2$ (according to Eq (4)). Then the E-step requires only $O(TC)$ time.

Regarding the M-step, it can be easily improved by a factor of $R^2$ (according to Eq (5)) again because each object is typically read in only a small number of locations. Next, we propose an optimization, called *candidate pruning*, to improve the M-step further to $O(TO)$, eliminating the factor of $C$. The idea is that in line 15, when computing the container that is most strongly co-located with a given object, it is probably safe to consider only containers that have been observed frequently with the object. So as a heuristic, we restrict the set of candidate containers to those that were most frequently co-located during the first several epochs. When testing for change points, we also include as candidates the most frequently co-located containers from recent epochs. Our experimental results show that candidate pruning is effective at reducing the time cost without affecting the accuracy.

We further propose a *memorization* technique that avoids unnecessary computation: If the set of objects in a container did not change in the previous EM iteration, then the location probabilities and co-location strengths for that container cannot change at the current iteration, so we can simply re-use the old values without any extra work. This optimization does not introduce any error.

Moreover, we can also use the history truncation method, described in Section 4 to truncate the history of size $T$, to a small critical region whose size is independent of $T$. All of the above optimizations combined finally reduce the **complexities of the E-step and the M-step** to $O(C)$ and $O(O)$, respectively.

### A.4 Extensions to the RFINFER Algorithm

*Missing container tags:* We have assumed that we know a priori which tags are container tags and which tags are object tags, which is typically true according to the EPC tag data standard[2]. Relaxing this assumption is possible. First, if the containers do not have tags, we simply omit the nodes in the graphical model that correspond to the readings from the container tags, along with the corresponding terms in Eq (4) and in line 6 of Algorithm 1. Alternatively, if all containers and objects have tags, but we do not know which are which, we treat all tags (for containers and objects) as if they were object tags; we use latent and evidence variables in the graphical model to denote the true and observed locations of these objects, respectively. Then, we use another copy of latent variables, $\ell_{tc}$, to represent real containers and use them to encode containment relationships with objects. Eq (4) and Algorithm 1 are modified the same as above. We do not further optimize these cases because they are rare given the wide adoption of the EPC tag data standard.

*Hierarchical containment:* Just as objects are grouped into containers, containers may themselves be stored in larger containers, such as pallets. We can extend our model and algorithms to arbitrarily nested containment hierarchies, intuitively by adding latent variables for the pallet locations whose values are imputed using EM in a similar way as the container locations. Since common types of queries such as those listed in Section 1 rely mainly on the immediate containers of items that are subject to shipping and packaging regulations, we defer a detailed study of hierarchical containment to future work when new applications requiring so emerge.

## B. DISCUSSION OF QUERY PROCESSING

We next describe our distributed query processing approach more.

First, we illustrate our approach using Query Q1 in Section 2. A monitoring query is split into local and global processing based on the labels of input streams of each query block. For instance, for Query 1, the inner query block takes two streams $R$ and $T$. Neither of them is labeled as "global", so this block is treated as local processing only. However, the output stream $S$ of the inner query block is labeled as "global", and hence the outer query block consuming $S$ is considered for global processing. For global query processing, the computation state is recognized and partitioned for individual objects. Revisit Q1. The outer query block employs pattern matching on the stream produced by the inner block. An automaton-based query processor [1] defines the query state to be: ($i$) the current automaton state, ($ii$) the minimum set of values needed for future automaton evaluation, e.g., the tag id and the time of its first exposure to room temperature for Q1, and ($iii$) the values that the query returns, e.g., the tag id and the sequence of temperature readings for Q1. Partitioning the query state for objects is simply based on the tag id. Then as an object leaves one site for another, we perform *state migration* by shipping the query state for this object to the next site or writing the query state into the tag's memory.

Second, we propose to exploit *stable containment* to share query states across objects. Our centroid-based sharing technique works as follows. Let $Q_o$ denote the query state for object $o$. We choose the most representative query state (the centroid) of all $Q_o$'s based on a distance function that counts the number of bytes that differ in the query state of two objects. The centroid selection problem has a $O(n^2)$ complexity, but since each case contains a limited number of objects, e.g., 20-50, this computation cost is modest on a modern computer. Given the centroid, we compress the query states of other objects based on the distance to the centroid.

## C. ADDITIONAL EXPERIMENTS

In this section, we describe additional experimental results beyond the key results presented in the main body of the paper.

---

[2]http://www.epcglobalinc.org/standards/tds/

Table 2: **Parameters used for generating RFID streams.**

| Parameter | Value(s) used |
| --- | --- |
| Number of warehouses ($N$) | 1 - 10 |
| Frequency of pallet injection (fixed) | 1 every 60 seconds |
| Cases per pallet (fixed) | 5 |
| Items per case (fixed) | 20 |
| Main read rate of readers ($RR$) | [0.6, 1], default 0.8 |
| Overlap rate for shelf readers ($OR$) | [0.2, 0.8], default 0.5 |
| Non-shelf reader frequency (fixed) | 1 every second |
| Shelf reader frequency (fixed) | 1 every 10 seconds |
| Frequency of anomalies ($FA$) | 1 every 10 - 120 seconds |

## C.1 Experimental Setup

We developed a simulator using CSIM to emulate an RFID-based enterprise supply chain. The parameters are shown in Table 2. Each supply chain arranges $N$ warehouses in a single-source directed acyclic graph (DAG). Pallets of cases are injected at the source, and then move through a sequence of warehouses with a scheduled delay in each warehouse and scheduled transit time between two warehouses, until they reach final destinations. Our simulator guarantees that in a period of time, pallets arrive at a warehouse and depart from it at the same rate (i.e., the system is in steady state).

Within a warehouse, pallets first arrive at the entry door and are read by the reader there. They are then unpacked. By default, each warehouse has a reader at the conveyor belt that scans the cases one at a time. The cases are then placed onto shelves and scanned by the shelf readers. After a period of stay, cases are removed from the shelves and repackaged. The assembled pallets are finally read at the exit door and dispatched to subsequent warehouses in a round-robin fashion. In the simulation, all readers have a read rate $RR$ for its location, uniformly sampled from [0.6, 1] unless stated otherwise. There is significant overlap between adjacent shelf readers: a shelf reader can read objects in a nearby location with probability $OR$ uniformly sampled from [0.2, 0.8]. Finally, to stress test our containment change detection algorithm, our simulator can inject anomalies that randomly pick an item and place it in a different case, with the frequency specified by the parameter $FA$.

Our evaluation uses the following metrics: **Error rate** (%): To measure accuracy, we compare the inference results with the ground truth and compute the error rate. **F-measure**: For change point detection, we evaluate the accuracy of the reported changes. We use *precision* to capture the percentage of reported changes that are consistent with the ground truth, and *recall* to capture the percentage of changes in the ground truth that are reported by our algorithm. We combine them into $F\text{-}measure = 2 * precision * recall/(precision + recall)$. **Running cost**: We report the time taken to evaluate a trace using a single-threaded implementation running on a server with an Intel Xeon 3GHz CPU and running Java HotSpot 64-bit server VM 1.6 with maximum heap size 1.5GB.

## C.2 Lab RFID Deployment

To evaluate our system in real-world settings, we developed an RFID lab with 2 ThingMagic Mercury5 readers connected to 7 circularly-polarized antennas, 20 cases containing 5 items each, and Alien squiggle Gen 2 Class 1 tags attached to all cases and items. We used the 7 antennas to implement 1 entry reader, 1 belt reader, 4 shelf readers, and 1 exit reader. Cases with contained items transitioned through the readers in that order, receiving 5 interrogations from each nonshelf reader and dozens from a shelf reader. The shelf readers had overlapping read ranges as they were placed close to each other. Using our lab setup, we created 8 traces with distinct characteristics, by varying the environmental noise, overlap among readers, and tag orientations:

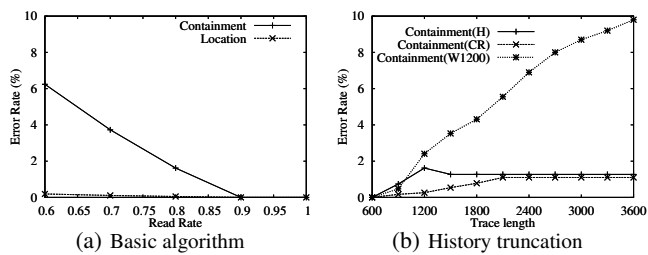

(a) Basic algorithm    (b) History truncation

Figure 6: **Experimental results for single-site inference**

- $T_1$ ($RR$=0.85, $OR$=0.25) represents the case of high read rates, an average of 0.85 across readers, and limited overlap rates, an average of 0.25 for shelf readers using low power.
- $T_2$ ($RR$=0.85, $OR$=0.5) is case of high read rates and significant overlap rates (using high power), an average of 0.5.
- $T_3$ ($RR$=0.7, $OR$=0.25) involves lower read rates due to added environmental noise, i.e., a metal bar placed on each shelf that is 1/3 the length of the shelf.
- $T_4$ ($RR$=0.7, $OR$=0.5) further has higher overlap rates.
- $T_5$ to $T_8$ extend $T_1$ to $T_4$, respectively, with containment changes. When all 20 cases were placed on shelves, 3 items were moved from one case to another and 1 item was simply removed, causing containment changes in 35% of the cases.

We also obtained traces with varied tag orientations but observed little impact of this factor. This verifies that squiggle tags are orientation-insensitive when used with circularly-polarized antennas.

## C.3 Alternative Method for Comparison

We describe the design of the SMURF$^*$ method used as a baseline for comparison in both our lab experiment and simulations. This method first uses SMURF [11] to smooth raw readings of objects to estimate their locations individually. The adaptive window used in SMURF is further stored for containment inference and change detection: Within the adaptive window for each item, at a particular time $t$, if the most frequently co-located case before time $t$ is the same as that after time $t$, then there is no containment change, and the most frequently co-located case is chosen to be the true container. Otherwise, we further check if none of the top-k co-located cases before time $t$ is in the set of top-k co-located cases after $t$. If so, we report a containment change for this item at time $t$, and pick the case that is most co-located with the item in the period from $t$ to the present. Note that the second check is needed because due to the missing readings, the real container may not be the most frequently co-located case, so the case that belongs to both top-k sets between and after $t$ could be the true container.

## C.4 Single-Site Inference

We describe several additional results for our inference methods when run on a single warehouse.

**Basic inference algorithm**. We first evaluate the basic algorithm presented in Section 3.2 for its sensitivity to various noise factors.

*Read rate*. We began with short 1500-second traces and ran inference with all the readings obtained thus far. We first varied the read rate $RR$ from 0.6 to 1. Figure 6(a) shows the inference error rates: Location inference is highly accurate, with the error rate less than 0.5% for all read rates. Containment inference is more sensitive to the read rate but still achieves an error rate less than 7% for the 0.6 read rate. The sensitivity of containment inference to the read rate is due to its use of co-location information: the chance of reading both an item and its container reduces quadratically with the read rate. Fortunately, the use of history alleviates the problem

336

**Table 3: F-measures (%) of containment change detection using different δ values and our offline method.**

|        | Threshold δ                                              |
|--------|----------------------------------------------------------|
|        | 10 | 20 | 30 | 40 | 50 | 60 | 70 | 80 | 90 | 100 |
| RR=0.6 | 64 | 71 | 75 | 85 | 89 | **89** | 87 | 87 | 87 | 87 |
| RR=0.7 | 85 | 88 | 90 | 92 | 94 | 96 | **94** | 93 | 91 | 90 |
| RR=0.8 | 92 | 97 | 98 | 97 | **97** | 97 | 96 | 95 | 93 |    |
| RR=0.9 | 83 | 98 | **97** | 97 | 97 | 97 | 97 | 96 | 96 | 95 |

and keeps the error rate low. The high accuracy of location inference is due to the effect of "smoothing over containment": once we understand containment correctly, the location of an object can be revealed by the readings of any other object(s) in the container.

*Overlap rate.* We also varied the overlap rate $OR$ from 0.2 to 0.8 while fixing the read rate at 0.7. Results show that neither location nor containment inference is sensitive to the overlap rate: the error rate of containment inference is flat at 2.3% and that of location inference is flat at 0.08%. This is because overall an object is read more by the reader closest to it than by other readers farther away.

*Container capacity.* We also varied the container capacity from 5 to 100 items while keeping the read rate $RR$ and overlap rate $OR$ at default values. The inference accuracy remains the same as the above with 20 items per container. This is mainly because for a given item, we calculate the weight of this item with its candidate container based on the co-location history; the weight calculation remains the same regardless of other items in the container. Hence, our inference algorithm is not sensitive to the container capacity.

**History truncation method**. In addition to the total inference time that we presented in Figure 5(b), we also computed the error rates of the three methods. Figure 6(b) shows the error rates of containment inference using the above methods (the error rates of location inference are always low in the 0.05% to 1% range, hence omitted). Again, we observe the naive window-based truncation to be inaccurate. Its error rate increases for longer traces because our simulation generates the belt readings useful for containment inference in the first half of the warehouse setup. In contrast, using the full history or the CR method makes inference not very sensitive to the trace length, with the CR method being somewhat better due to the elimination of noisy data from inference.

**Containment change detection**. We first examine the effect of the threshold δ for change point detection. We consider fixed values in a range as well as our offline method as described in Section 3.3. We created traces with varied read rates. Table 3 shows the *F-measure* for these traces as δ takes various fixed values. The bold numbers indicate the F-measures of the δ chosen by our offline sampling algorithm. As can be seen, the best fixed threshold that gives the optimal F-measure varies across traces. Our chosen threshold always approximates the optimal value within 2% across all read rates.

We further study the tradeoff between accuracy and efficiency. The change point detection algorithm requires a recent history (whose size is $\bar{H}$), besides the critical region in the past, to detect containment changes. Since we run inference with the default frequency of once every 300 seconds, $\bar{H}$ has to be at least 300 seconds to include all new observations for inference. However, a longer recent history may be needed to ensure accuracy of change point detection while it may also increase inference cost. We next vary $\bar{H}$ to study such tradeoff between accuracy and efficiency. We created traces with varied read rates and for each trace, measured F-measures and time costs as different $\bar{H}$ values were used.

Table 4 shows the results. The overall trend with all traces is that as $\bar{H}$ increases, the F-measure improves but the time cost also increases. Among different read rates, we see that achieving high accuracy for lower read rates such as 0.6 requires larger sizes of $\bar{H}$, while the time cost varies with $\bar{H}$ consistently across all read rates. This implies two ways to trade off accuracy and efficiency: (1) If the application requirement is to achieve best accuracy while keeping up with stream speed, we should use the $\bar{H}$ sizes that yield the time costs in bold in the table because they allow inference to complete within 300 seconds, the interval before the next inference starts. This way, our algorithm offers above 90% accuracy for read rates ≥ 0.7 and above 80% for the read rate = 0.6. (2) If the requirement is to optimize performance while satisfying certain accuracy, say, 85%, we should use the $\bar{H}$ sizes that yield the F-measures in bold in the table. In particular, when the read rate is high, we can use a smaller size of $\bar{H}$ to reduce the time cost and enable more frequent inference (e.g., for read rate=0.8, running inference every 212 seconds can keep up with the stream speed). Overall, for the common read rates between 0.7 and 0.9, the $\bar{H}$ size of 500 seconds achieves over 90% accuracy while running at stream speed.

**Table 4: F-measures (%) and time costs (sec) of change point detection with different recent history sizes ($\bar{H}$) for different read rates ($RR$).**

|        |          | Recent history size $\bar{H}$              |
|--------|----------|---|---|---|---|---|---|---|
|        |          | 300 | 400 | 500 | 600 | 700 | 800 | 900 |
| RR=0.6 | F-m. (%) | 46 | 67 | 81 | **87** | 88 | 86 | 92 |
|        | Time (s) | 205 | 235 | **293** | 385 | 418 | 497 | 556 |
| RR=0.7 | F-m. (%) | 72 | **91** | 90 | 90 | 93 | 93 | 94 |
|        | Time (s) | 182 | 229 | **288** | 344 | 403 | 469 | 523 |
| RR=0.8 | F-m. (%) | 73 | **92** | 94 | 97 | 93 | 95 | 96 |
|        | Time (s) | 182 | 220 | **283** | 341 | 395 | 446 | 490 |
| RR=0.9 | F-m (%)  | 82 | **94** | 95 | 98 | 93 | 95 | 96 |
|        | Time (s) | 172 | 207 | **258** | 322 | 381 | 436 | 458 |

**Summary**: ($i$) Our inference algorithm is highly accurate for various noisy traces with stable containment (≤7% error rate for containment inference and around 0.5% for location inference). The critical region method can significantly reduce inference cost for long traces while further improving accuracy. ($ii$) Our results are stable for various read rates, overlap rates between readers, container capacities, and history lengths. With change point detection, our algorithm can achieve 85% accuracy given read rates ≥ 0.7 while keeping up with stream speed (by using a relatively small recent history size). These results are confirmed using both real lab traces with various noise factors and simulations with different containment change frequencies.

### C.5 Distributed Inference

We highlight the communication costs of the distributed and centralized approaches in the table below (while omitting other results in the interest of space). For the centralized approach, we assume that all raw data is shipped to a central location for inference with simple gzip compression of data. As can be seen, our CR methods offer 3 orders of magnitude reduction in communication costs.

**Table 5: Communication costs (bytes) of a centralized approach and three state migration methods for distributed inference.**

|        | Centralized | None | CR      |
|--------|-------------|------|---------|
| RR=0.6 | 125,895,500 | 0    | 225,890 |
| RR=0.7 | 145,858,950 | 0    | 223,790 |
| RR=0.8 | 166,746,235 | 0    | 225,890 |
| RR=0.9 | 187,589,810 | 0    | 225,890 |

**Summary:** Our results show that distributed inference using the CR methods have accuracy that is close to the centralized approach, while incurring significantly lower communication costs. Our scalability results (in Section 5.3) show that our distributed inference system can scale to millions of objects over multiple sites while keeping up with the speed of RFID streams at each site.